# Near-single-photon atto-watt detection at mid-infrared wavelengths by a room-temperature balanced heterodyne set-up

Daniele Palaferri[1,a], Lorenzo Mancini[1], Chiara Vecchi[1], Leonardo Daga[1], Pierfrancesco Ulpiani[2], Massimiliano Proietti[2], Carlo Liorni[2], Massimiliano Dispenza[2], Francesco Cappelli[3], Paolo De Natale[3], Simone Borri[3]

1. Photonics Research & Applied Navigation Science Lab, GEM elettronica S.R.L., Via Amerigo Vespucci, 9 63074 San Benedetto del Tronto (AP) – ITALY

2. Leonardo Innovation Labs, Via Tiburtina km 12400, Rome, 00131, Italy,

3. CNR-INO - Istituto Nazionale di Ottica, Largo E. Fermi 6, 50125 Firenze – Italy

a. E-mail: daniele.palaferri@gemrad.com

**Abstract**

**Single photon detection is the underpinning technology for quantum communication and quantum sensing applications. At visible and near-infrared wavelengths, single-photon-detectors (SPDs) underwent a significant development in the past two decades, with the commercialization of SPADs and superconducting detectors. At longer wavelengths, in the mid-infrared range (4-11µm), given the reduced scattering and favourable transparent atmospheric windows, there is an interest in developing quantum earth-satellites-links and quantum imaging for noisy environments or large-distance telescopes. Still, SPD-level mid-infrared devices have been rarely reported in the state-of-the-art (superconductors, single-electron-transistors or avalanche-photodiodes) and, crucially, all operating at cryogenic temperatures. Here, we demonstrate a room-temperature detection system operating at 4.6µm-wavelength with a sensitivity-level of atto-watt optical power, corresponding to few tens of mid-infrared photons. This result was obtained by exploiting a pair of commercially available photodetectors within two balanced-heterodyne-detection setups: one involving a quantum-cascade-laser (QCL) and an acousto-optic-modulator (AOM) and the other one including two QCLs with mutual coherence ensured by a phase-lock-loop (PLL). Our work not only validates a viable method to detect ultra-low-intensity signals, but is also potentially scalable to the entire wavelength range already accessible by mature QCL technology, unfolding - for the first time - quantum applications at mid- and long-wave-infrared-radiation.**

Photons in the mid-infrared range have energies in the order of $10^{-20}$ J, *i.e.* few percentages of 1 atto-joule ($1\times10^{-18}$ J). The capability of detecting these low energy values requires extremely sensitive devices. In order to quantify the detection sensitivity, an important figure of merit is the Noise-Equivalent-Power (NEP), which is defined as the minimum detectable signal power to achieve signal-to-noise-ratio (SNR) = 1 within 1Hz bandwidth. For classical photo-detectors in the shot-noise limit, NEP is expressed as NEP = $\Delta f\, h\nu/\eta$ [1], where h is the Planck constant, ν the radiation frequency, η the device absorption efficiency and Δf is the temporal bandwidth. For SPDs the expression takes into account the dark-count-rate (DCR)[2], i.e.: NEP = $h\nu(2\cdot DCR)^{0.5}/\eta$. At visible and telecom wavelengths, conventional SPDs consist of photomultipliers and avalanche-photodiodes[3] based on silicon or indium-gallium-arsenide materials, whose NEP is in the range of 1 atto-watt ($1\times10^{-18}$ W) [4]. In the mid-infrared range, commercially available best detector devices – consisting of small-band-gap photodiodes based on mercury-cadmium-telluride (MCT) alloys[5] or super-lattice III-V detectors such as quantum-well-infrared-photodetectors (QWIPs)[6] and quantum-cascade-detectors (QCDs)[7] – are still far from achieving SPD-level performances: indeed, the typical reported NEP for these devices is in the order of the pico-watt ($1\times10^{-12}$W) power level and a corresponding detectivity in the range $10^{10}$-$10^{11}$ cm·Hz$^{1/2}$/W [8]. In the academic literature, to the best of our knowledge, three noteworthy mid-infrared devices have reported

SPD-level sensitivity: 1) avalanche-photodiodes made of MCT material[9] operating in the wavelength-range 0.9-4.3 μm and at temperatures 78K-110K; 2) tungsten-silicide superconducting detectors[10] operating in the wavelength-range 4.8-10μm and at temperatures below 4K ; 3) charge-sensitive-infrared-phototransistors (CSIPs)[11,12], based on III-V heterostructures and a single-electron-transistor concept, demonstrated in the long-wavelength-range 9-50μm[13] and operating with atto-watt sensitivities at liquid He temperatures (4-30K)[12] . All these devices, with remarkable SPD performances and some already envisaged for single-molecule spectroscopy[14], are currently not compatible with room-temperature operation, restricting their suitability in future real-world applications.

An alternative technique to detect ultra-weak fields down to single photons is the optical heterodyne interferometry technique[15,16], which involves the mixing of a powerful local oscillator (LO) laser with a weak signal (S) laser onto a fast detector. Typically, this scheme, in its balanced version, is implemented by superimposing two laser beams through a beam-splitter component: the resulting beat-note signal, falling in the radiofrequency (RF) range, is strictly related to the weak signal S of interest (see sketch in **Fig. 1a**) and can be easily acquired, processed and analysed with standard RF equipment. Heterodyne interferometry - widely popular in quantum optics[17] - has been successfully employed in various fields requiring extreme performances, such as astronomical imaging and spectroscopy[18], gravitational-waves detectors[19,20], quantum detection for continuous-variable quantum-key-distribution[21] or optical tomography techniques to reconstruct photon-number statistics of various sources[16].

Recently, fast heterodyne receivers have been demonstrated at mid-infrared wavelengths by exploiting the intrinsic short lifetime of intersubband transition in III-V QWIPs and QCDs with reported heterodyne-NEP values of 200fW and 2pW, respectively[22] and a large beat-note response up to 110 GHz[23]. Preliminary studies have been carried out to employ these devices as an astronomical photonic correlator for distant telescopes imaging[24], receivers for free-space-optical communication, frequency-modulated-continuous-wave ranging[25,26] and photo-mixers for the generation of sub-THz radiation[27].

In the present work, at first, we show theoretically and experimentally that, at 4.6 μm-wavelength, the use of an RF-stabilised balanced heterodyne detection (BHD) setup with commercial MIR photodetectors allows to obtain performances quite close to standard SPD receivers, with NEP values in the atto-watt range. Then, we discuss how these results are exploitable in photonic integrated circuit platforms and in few-photons interferometry schemes for innovative quantum sensing applications at mid-infrared wavelengths.

**Results**

In a BHD scheme, the signal-to-noise ratio corresponds to the ratio between the heterodyne photocurrent $i_{het}$ and detection system current noise $i_n$ at the beat-note frequency $\omega_h$, which are expressed as[1]:

1) $i_{het}(\omega_h) = 2R\sqrt{P_{LO}P_S(\omega_h)}$
2) $i_n(\omega_h) = \sqrt{\Delta f \cdot \Sigma_j S_j(\omega_h)} = \sqrt{\Delta f \cdot [S_{det}(\omega_h) + S_{LO}(\omega_h)]}$

Where R is the detector Responsivity, $P_{LO}$ is the LO power, $P_S$ is the signal power, $\Delta f$ is the detection bandwidth, the beat-note frequency ($\omega_h = \omega_{LO} - \omega_S$) is the spectral difference between the peak position of the LO and the signal (we are assuming perfectly mono-chromatic optical beams), $\Sigma_j$ is a sum on all $j$ noise components and $S_j$ is the Power Spectral Density (PSD) component which is measured in ampere-square-per -hertz $\left(\frac{A^2}{Hz}\right)$. $S_{det} = S_d + S_{th} + S_{1/f}$ is the intrinsic PSD of the detector (containing dark current noise $S_d$, thermal noise $S_{th}$, 1/f noise $S_{1/f}$), and $S_{LO} = S_{shot} + S_{lfn} + S_{rin}$ is the PSD induced by the LO-laser (containing the shot-noise $S_{shot}$, the laser-frequency-noise $S_{lfn}$ and the relative-intensity-noise $S_{rin}$). For the analytical expression of these terms and a description of all PSD figures involved in a BHD system, see Supplementary Information.

The heterodyne-NEP is obtained retrieving $P_S$ at the condition $\left[\frac{S}{N} = 1\right]$, i.e., 1) = 2):

3) $\quad \text{NEP}_h = \frac{\Delta f \cdot \Sigma_j S_j}{4R^2 P_{LO}} \approx \Delta f \cdot \frac{S_{det} + S_{shot}}{4R^2 P_{LO}} = \Delta f \cdot \left(\frac{S_{det}}{4R^2 P_{LO}} + \frac{eg}{2R}\right)$

Where we assume $S_{LFN}, S_{RIN} \ll S_{shot}$ and we use the expression $S_{shot} = 2egRP_{LO}$ including the electron charge e and the photocurrent gain g. A sufficiently strong LO field allows to overcome intrinsic detector noise sources – which limit MIR photodetector performances at room-temperature - and to reach a shot-noise-limited heterodyne-NEP ultimately dependent only on the photodetector responsivity $R = egh\nu/\eta$ and therefore on the photon energy $h\nu$ and the quantum efficiency $\eta$[1]: $\text{NEP}_{h-SN} = \Delta f \frac{eg}{2R} = \Delta f \frac{h\nu}{2\eta}$.

In **Table 1**, we compare standard MIR detectors that are either commercially available or currently investigated in academic labs: a QCD in a ridge-waveguide configuration[28], a QWIP in 45°-facet configuration[29] and an MCT[30]: these three devices, with large frequency cut-offs (0.5-20GHz), are all suitable as heterodyne receivers with QCL pumping sources. Looking at the linear-NEP values (taken from datasheets) and the heterodyne-NEP values (computed by using the above definition of NEP$_{h-SN}$), we notice that the heterodyne technique has the potential to enhance up to ~8 orders of magnitude their photodetection performance.

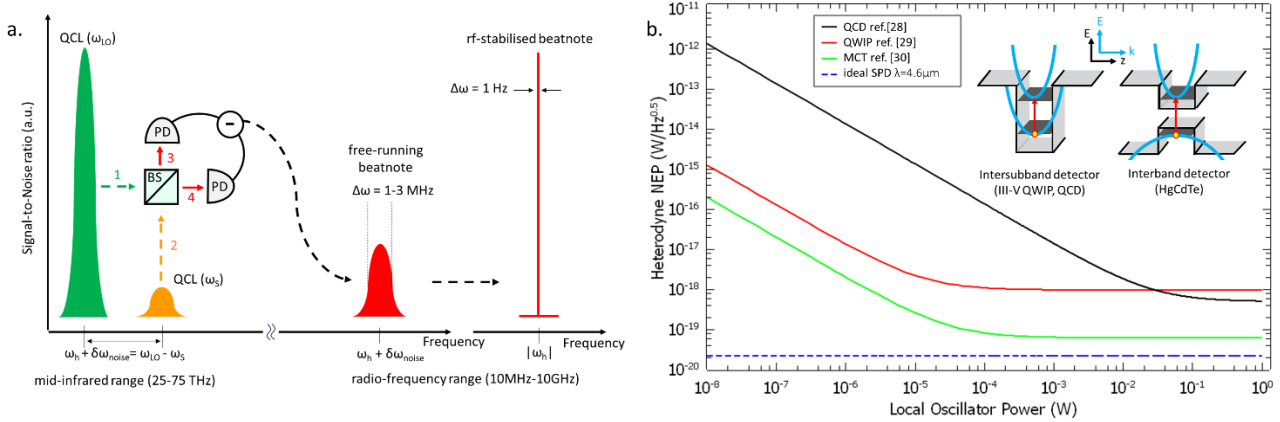

**Figure 1| BHD detection at mid-infrared wavelengths and heterodyne noise-equivalent-power performances. a,** Ideal BHD scheme involving QCL sources (one as LO and one as weak intensity signal S), a free-space beam-splitter (BS) and fast photodetectors (PD) at the ports (3) and (4); the retrieved rf-beatnote is illustrated in its free-running-laser linewidth and in a rf-stabilised signal-linewidth; **b,** heterodyne noise-equivalent-power for currently available MIR detectors; the dashed blue line shows the $\text{NEP}_{h-SN}$ for an ideal detector with $\eta = 100\%$; the inset on the top shows a sketch of super-lattice band structure (energy E vs growth axis z and vs k-vector) of intersubband detectors (QWIPs[29] and QCDs[28]) and interband detectors made of HgCdTe[30], with the red arrow indicating the electronic transition in each structure.

The two regimes of equations (2) and (3) are illustrated respectively in **Fig. S2** and **Fig. 1b**, showing the current noise and the NEP$_h$ as function of the LO power. In the detector-limited regime, the current noise is constant and the dominating term is $S_{det}$ (dark current and thermal noise) while in the shot-noise-limited regime the LO power is the only contribution and the current noise scales with $(P_{LO})^{1/2}$. Analogously, the left term of equation (3) is progressively reduced as $P_{LO}$ increases, until the NEP$_h$ reaches a constant term dependent only on detector responsivity. Interestingly, in the shot-noise regime it is not necessary to increase further the LO power to amplify the detector performances. This result is particularly important because it lessens the high-power requirement for QCL sources and allows high performances in MCT detectors, which have limited power saturation intensity[30], of the order of 1 W/cm$^2$. QWIPs and QCDs, at the same time, have much higher saturation intensity (~10$^7$ W/cm$^2$)[31] thanks to the ~ps intersubband transition lifetime, but their lower responsivity limits the NEP$_h$ to higher values with respect to MCT devices.

The NEP$_h$ values reported in **Table 1** and in the shot-noise-limit region of **Fig. 1b** - comparable to SPD performances - are difficult to obtain experimentally because they correspond to an ideal case with a lossless beam-splitter (see **Fig. 1a**)[32] and laser sources treated as Dirac delta functions. In real-world conditions, a free-space heterodyne setup is made of lossy beam-splitters and lenses (made of calcium fluoride or zinc selenide crystals and other coatings) and metallic mirrors: all these optical components add absorption losses, non-zero phase delays and optical unbalancing among transmission/reflection light paths. Also, free-running DFB QCL sources are limited by their frequency-noise[33] and have a typical linewidth of 1-3 MHz[34] .

Therefore, three actions are required to achieve high-performance detection in a mid-infrared BHD: 1. developing a sensitive and high-bandwidth detection system, characterized by fast and ultra-low-noise electronics, allowing to select the optimal range for heterodyne detection; 2. optimizing the balanced configuration in order to completely suppress the photocurrent DC terms generated by the LO laser and additional noise figures coming from the optical and electrical arrangements; 3. providing an RF-stabilization technique to remove the reciprocal frequency fluctuations between the two free-running QCLs, which is translated into intensity noise by the BHD (as depicted in the sketch of **Figure 1a**).

**Fig. 2** illustrates the two approaches we have investigated to achieve the targeted detection performance. The first one (**Fig. 2a**) includes a DFB QCL emitting at λ = 4.6 µm, an acousto-optic modulator (AOM) that induces a first-order spectral shift of $f_{AOM}$ =105 MHz, various mid-infrared optical components and a detection module based on a couple of identical MCT photodetectors and on custom-made low-noise and high-bandwidth electronics. The beating between the LO main beam and the AOM-shifted beam (representing the signal S) provides a stable heterodyne signal at $f_{AOM}$. The second setup (**Fig. 2b**) involves two different DFB QCLs - one acting as LO and one as S - which are both emitting at 4.6 µm-wavelength and are spectrally matched by optimal current/temperature points, the same balanced detection module, a third detector (MCT with 200 MHz bandwidth) to provide the beat-note reference for a commercial electronic phase-locked-loop (PLL) and various beam-splitters to superpose the optical beams: in this case the stabilised BHD signal is exactly at the clock frequency $f_{clock}$ set by the PLL to correct the current variations of the LO QCL. In both the setups the BHD signals are amplified by commercial RF-amplifiers and acquired by an electrical spectrum analyser.

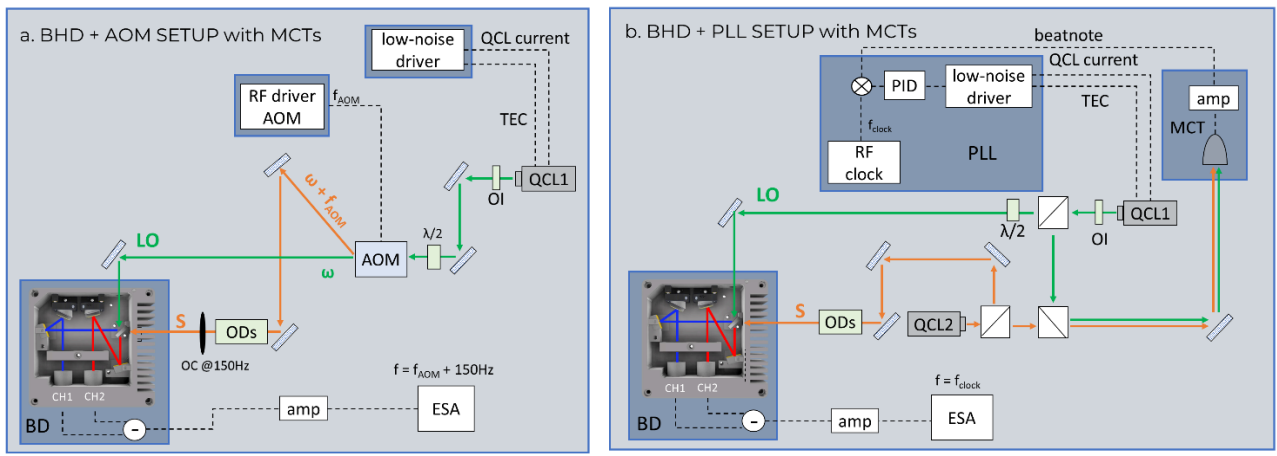

**Figure 2 | Experimental BHD setups with MCT detectors**. See main text and Methods for operational description. A low-noise laser driver controls DC current and temperature of LO QCL (green arrows), TEC: thermal electric cooler, OI: optical isolator, AOM: acoustic-optical modulator, ω: optical frequency of QCL, $ω_{AOM}$: frequency shift induced by AOM generating the S beam (orange arrows), ODs: optical density filters, OC: optical chopper placed in front of the S beam, BD: balanced detectors module with an internal calcium fluoride beam-splitter, output the current difference between the two MCT detectors at channel 1 (CH1) and channel 2 (CH2), amp: RF-amplifier, ESA: the electrical spectrum analyser is used for beat-note signal characterization. In the BHD+PLL setup, a third MCT detector is used to provide the feedback for the PLL module, which is mixed with an RF clock signal.

Heterodyne spectra are collected by progressively reducing the power of the signal, through interferometric optical density filters (as in [22]), in the following conditions: free-running single-heterodyne-detection (SHD) with two QCLs and only one MCT with 300 kHz resolution bandwidth (**Fig. S3a**), free-running BHD with two QCLs and the balanced module with two MCTs with 300 kHz resolution bandwidth (**Fig. S3b**), RF-stabilised BHD with AOM-based and PLL-based setups (**Figs. 2a** and **2b**) with 1Hz resolution bandwidth. We emphasize that the stabilised approaches allow to lower the resolution bandwidth down to 1Hz, disclosing detection performances that are various orders of magnitude beyond the datasheet figures of merit of the photodetectors. All the results are reported in the graph **Fig. 3a** (heterodyne-power-signal in dBm vs optical power signal). We can notice that a SHD scheme without stabilization, allows to measure a minimum signal Ps = 300 fW (black points) which is one order of magnitude lower than the linear NEP = 5pW of the commercial MCT device (see **Table 1** e ref.[30]) and comparable to state-of-the-art results reported elsewhere[22]. By implementing a BHD scheme without stabilization, it was possible to detect a minimum heterodyne signal Ps = 30 fW (red points), *i.e.* 10-fold enhancement in detection performances with respect to the SHD case thanks to the parasitic noise suppression. Finally, by employing the RF-stabilised schemes of Figure 2, we have measured Ps = 10aW with the BHD+PLL setup and Ps = 1.4 aW with the BHD+AOM setup which are respectively three and four orders of magnitude better than the data obtained with the BHD scheme without stabilization. These unprecedented results were obtained by carefully optimising the optical alignment for the balanced detection signal and by distancing few-meters away the laser drivers and the RF-clocks from the detection module to reduce electrical noise interference. The BHD+AOM set of experimental data follows the theoretical BHD curve (pink line in the graph) which is defined as $I_h = 4ZR^2P_SP_{LO}G_{AMP}$, where $I_h$ is the heterodyne power signal (as it is measured on the spectrum analyser), $Z = 50\Omega$ is the input impedance, R is the detector responsivity and $G_{AMP}$ is the total amplifiers gain: the non-linearity of the experimental points at the lowest signal intensities (1-10 aW) is caused by the increasing noise contribution (dashed lines) to the peak signal measurement. While the experimental data are still higher than the shot-noise quantum limit computed with equation (3) ($P_S$ = 0.06 aW), the possibility to measure at room-temperature optical powers in the atto-watt range with 1Hz integration - corresponding to the detection of ~20 photons per second - represents a milestone for mid-infrared technologies. Indeed, by comparing the data with the linear NEP term $I_l = ZR^2(NEP_l)^2G_{AMP}$ – where we use the linear NEP from detector datasheet - we have enhanced the detection sensitivity of the MCT detector by more than six orders of magnitude. Our setup could be further improved by employing proper low-noise RF-amplifiers (to minimise their Johnson-Nyquist noise contribution) and using a digital data acquisition board as it is typically implemented with SPD systems (to sample a high number of data and to provide suitable statistics with respect to the dark-count-rate)[15]. Moreover, the other three sets of experimental data ('BHD+PLL', 'SHD' and 'BHD') show lower heterodyne intensity signal with respect to the experiment with AOM: this happens because in the unbalanced experiments a large portion of the signal is distributed among the ~MHz linewidth of the beat-note while in the experiment with the PLL the electronic feedback loop is not as efficient to stabilise the beat-note as it occurs with the experiment involving the AOM. The heterodyne signal spectra of the BHD+AOM setup are reported in **Fig. 3b**: remarkably, at 1 femto-watt input $P_S$ power there is a 25dB BHD signal and at 1 atto-watt there is evident a few dBs signal above the ground noise of the electrical spectrum analyser. **Fig. 3c** shows the spectra of the BHD+PLL setup, where the 10 atto-watt signal is quite evident and the noise spectra contain a residual peak at -95dBm due to the RF-clock: this noise is currently limiting the BHD+PLL setup with MCTs and its suppression could allow to obtain similar performances to the BHD+AOM setup.

Similar experiments with analogous conditions have been carried out with a couple of identical QCDs with responsivity peak at 4.6 μm[28] – see **Fig.s S3, S4** and **3d**: while the SHD and BHD systems with free-running QCLs reached respectively 1 nW and 100 pW as minimum detected signal, the BHD-PLL setup has obtained 2 fW – showing five orders of magnitude improvement. **Fig. 3e** shows the beat-note spectra with PLL at 1Hz resolution bandwidth, with the 2 fW signal being few dBs above noise. The heterodyne spectra acquired in the other conditions of **Fig. 3a** and **Fig. 3d** are reported in the Supplementary materials (**Fig. S5**). The results with commercial QCDs did not show the atto-watt range detection performance, as it occurred with the MCT detectors. This is ascribed to the fact that QCDs responsivity – crucial for the $NEP_{h-SN}$ – is three orders of magnitude smaller than the responsivity of MCTs (see **Table 1**); moreover, these commercial QCDs are realised in a ridge-cavity configuration – similar to QCLs – which induces optical reflections and parasitic locking to Fabry-Perot resonances, increasing the detection noise and lowering the PLL efficiency.

**Discussion**

We believe that III-V intersubband detectors could reach MCT performances by employing recently reported high-responsivity and low-noise configurations, such as a patch-antenna array architecture[35] or optimised structures with single-period QCDs[36]. Moreover, the advantage of using QCD detectors relies not only on their high-speed but also on their potential exploitation combining a unique bi-functional structure QCL/QCD[37] and InP-based integrated photonics. Superlattice structures can indeed be conveniently integrated by a homogenous process with epitaxial InGaAs waveguides on InP substrate[38] : thanks to Fe-doping-based growth and optimised dry-/wet-etch processes, these waveguides have been recently reported with low-loss performances (0.5-1 dB/cm) in the MIR range 5-11 µm[39,40]. **Fig. 4a** illustrates the concept of a photonic integrated circuit (PIC) including the heterodyne scheme with DFB QCLs and QCDs. The main advantages of the integrated platform consist of the intrinsically lower loss than the bulk setup (a PIC would show ~0.1 dB signal attenuation in few-millimetres waveguide path while the free-space setup is limited by the 1-3dB transmission losses of standard MIR optical components) and the convenient fabrication yield of a monolithic PIC with high-reproducibility (InP photonics has reached a random defect density wafer <1 cm$^{-2}$ in the last decade[41]) and extreme compactness ( ~1cm$^2$ chip-size as compared with ~1m$^2$ dimensions of the free-space setups developed in this work). Also, the InP-chip of BHD could be embedded into an application-specific integrated circuit (ASIC) including the laser electronics (the low-noise drivers, the mixer and the clock for the PLL) and the read-out components (the amplifiers, the analog-digital converters and the FPGA to sample the data) – as already demonstrated with InGaAs/InP-based SPAD sensors[42] .

Finally, to take advantage of the high performances of our BHD system, we have performed interferometric measurements at low intensity level as shown in the experimental setup of **Fig. 4b** and the results in **Fig. 4c**. By inserting a piezo-electric mirror in the *signal* path of the BHD-PLL setup (see figure 4b), we have acquired the heterodyne signal as a function of the piezo position. As we increase the number of optical density filters, attenuating the signal from OD7 (100 pW) to OD13 (100 aW), we notice that all attenuated signals follow the $[1+\cos x]^2$ dependency, as expected in a heterodyne interferometry scheme (see Supplementary Materials). Remarkably, even the 100 aW signal has an extinction ratio of ~15 dB. Moreover, all data have a periodicity of 2.1±0.2 µm, quite close to λ/2 (deviations may be caused by piezo instabilities). Being able to handle the phase of the signal field allows to implement a few-photons interferometry system which could be the core of various applications where high sensitivity is required: a target at very large distances or with poor reflectivity in a ranging or quantum imaging experiment[43], a transmission channel for free-space optical communication in a noisy environment[25], a sample with single-molecules to be identified[14]. Moreover, the phase shift of the piezo-electric mirror could be replaced in a PIC by thermo-optic modulators or superlattice-based Stark modulators[44].

In conclusion, we have demonstrated atto-watt signal sensitivity at 4.6 µm wavelength by developing a room-temperature RF-stabilised low-noise and high-bandwidth BHD system. Further improvements towards single-photon-detection could be obtained by optimising the signal acquisition or replacing the LO QCL with a stable low-noise frequency-comb-QCL source (as recently demonstrated in [45]). A PIC-based MIR single-photon sensor operating in the mid-infrared could be potentially installed in nano-satellites or unmanned-aerial-vehicles[46] for future payload-missions regarding exoplanet investigation[47], atmospheric remote sensing[48] or high-bit data-rate quantum encrypted transmission[49].

Table 1 Comparison of parameters of currently accessible fast mid-infrared detectors: QCD[28], QWIP[29] and MCT[30].

| Operational parameters and figures of merit | QCD[28] | QWIP[29] | MCT[30] |
|---|---|---|---|
| Peak Wavelength $\lambda_p$ [μm] | 4.65 | 4.95 | 4.70 |
| Detector Temperature $T_{det}$ [K] | 295 | 295 | 200 |
| Quantum Efficiency $\eta$ [%] | 2.4 | 4.2 | 34.4 |
| Photoconversion Gain g | 0.01 | 0.75 | 1 |
| Responsivity R [A/W] | $1.8 \times 10^{-3}$ | 0.125 | 1.3 |
| Specific Detectivity D* [cm·Hz$^{1/2}$/W] | $1.5 \times 10^{9}$ | $7.0 \times 10^{7}$ | $2.0 \times 10^{10}$ |
| Frequency-cut-off fc [GHz] | 20 | 26 | 0.5 |
| Electrical Area $A_e$ [mm$^2$] | ~$2.5 \times 10^{-3}$ | $9.0 \times 10^{-4}$ | 1.0 |
| Linear NEP [W/Hz$^{1/2}$] | $3.0 \times 10^{-10}$ | $4.0 \times 10^{-11}$ | $5.0 \times 10^{-12}$ |
| Heterodyne NEP$_{h-SN}$ [W/Hz$^{1/2}$] | $5.1 \times 10^{-19}$ | $9.6 \times 10^{-19}$ | $6.2 \times 10^{-20}$ |

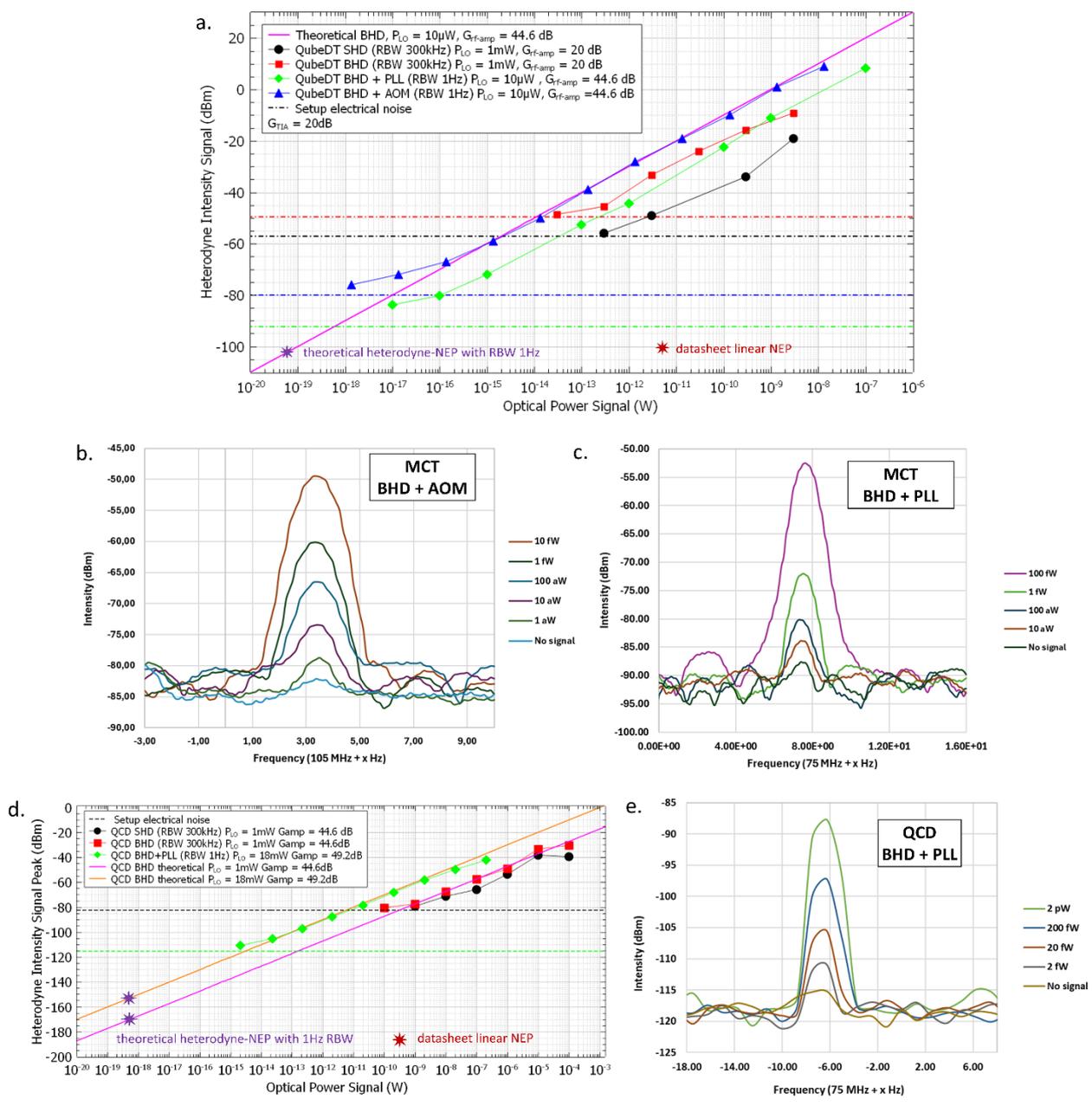

**Figure 3 | Heterodyne detection performance measurements. a.** Heterodyne detection peak signal (dBm) as function of the optical power signal of QCL S (W) for BHD setups with MCT detectors, acquired at various resolution bandwidth (RBW). Points data are measured on the electrical spectrum analyser, dashed lines indicate ground noise without optical signal, the continuous pink line is a theoretical curve, the purple and brown stars indicate respectively the heterodyne-NEP point (analytically computed detector quantum limit with RBW 1Hz - see main text) and the linear NEP point (from detector datasheet); **b.** Heterodyne spectra of BHD+AOM setup with MCT detectors (RBW 1 Hz); **c.** Heterodyne spectra of BHD+PLL setup with MCT detectors (RBW 1 Hz); **d.** Heterodyne detection peak signal (dBm) as function of the optical power signal of QCL S (W) for BHD setups with QCD detectors; **e.** Heterodyne spectra of BHD+PLL setup with QCD detectors (RBW 1 Hz).

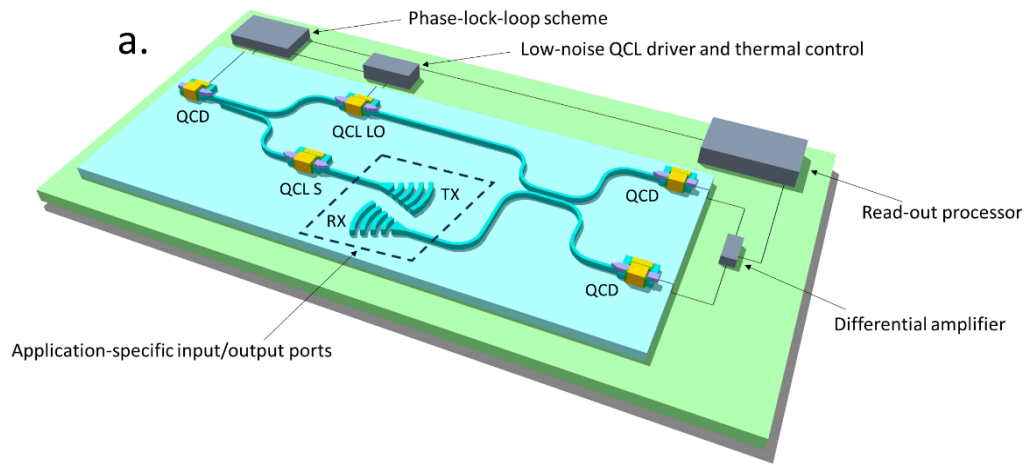

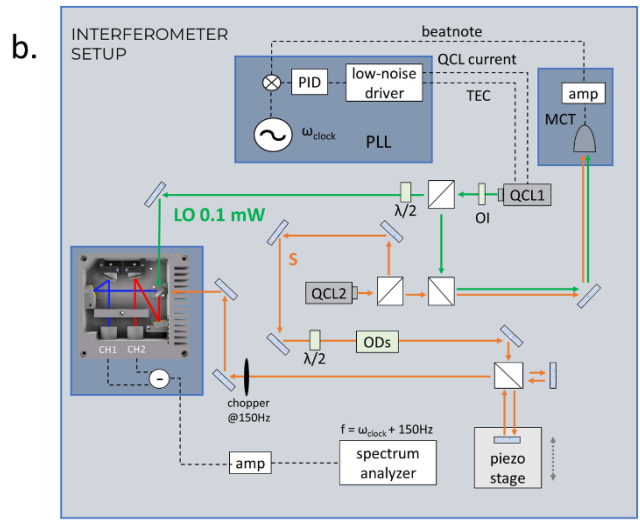
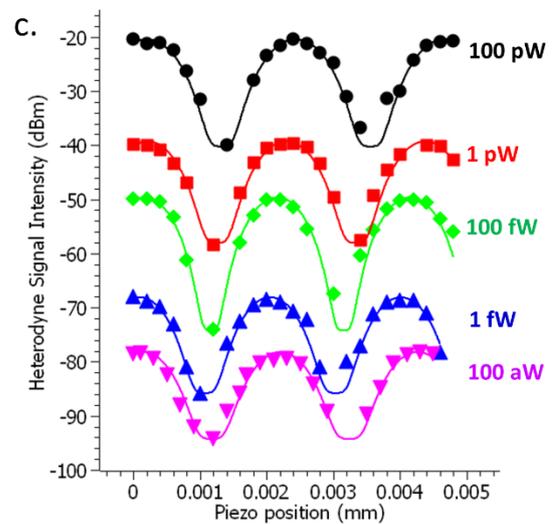

**Figura 4 – BHD PIC concept and few-photons heterodyne interferometry. a.** Conceptual illustration of integrated single photon heterodyne detector based on InP monolithic photonic integrated circuits with DFB QCL acting as LO and S lasers and three QCDs (two for the balanced detection and a single one for the phase-lock). The S beam could be out-coupled and back-coupled through the use of grating couplers (TX: transmitter port, RX: receiver port) depending on the specific target application; **b.** Setup with PLL+BHD for heterodyne interferometry measurements employing a beam-splitter and a piezo-electric stage; **c.** Experimental results at various signal powers in the range 100pW – 100aW, all data follow $(1+\cos x)^2$ fit.

## Online Methods

The balanced MCT detectors are PVI-2TE5 from Vigo Photonics embedded in the Qube-DT module from PpqSense S.R.L. The QubeDT has an AC transimpedance amplifier $r_{TIA} =10^4$ Ohm. The QCDs are from Hamamatsu and are balanced by an electronic 180° power splitter. DFB QCL sources are provided by Thorlabs. Laser drivers and the PLL module are from PpqSense S.R.L.: in all setups the QCL drivers have a reduced current noise (~100pA/Hz$^{0.5}$) in order to minimise the lasers' frequency-noise. For all experiments the LO power is kept in the range ~10µW-1mW for MCTs and 1-20mW for QCDs to minimise electrical noise of the setup (depending on the configuration). The beat-note RF-amplifiers are ZFL-500 and ZX60 from Mini-Circuits with gain respectively 24.6dB and 20dB in the range 70-150 MHz: in some configuration only one amplifier was used, while in other two of them in cascade were employed to amplify the beatnote signal; increasing amplifier gain is beneficial only to overcome the instrumentation noise (ground noise of the spectrum analyser), however detector noise is amplified as much as the signal, therefore to detect low optical intensities it is not beneficial to increase further the number of amplifiers. For the BHD+AOM setup with MCTs (Figure 3a) and the BHD+PLL setup with QCDs (Figure S4), at the beat-note signal frequency there is an important pick-up noise emitted at the same electronic frequency, respectively by the AOM RF-driver (at $f_{AOM}$) and by the RF-clock (at $f_{clock}$): these noise signals are indistinguishable from the heterodyne beat-note; therefore, to discriminate the BHD signal, the S beam is modulated by means of an optical chopper spinning at 150Hz and we calibrate the system detection performances at the first sideband frequency i.e., $f = f_{AOM}$ + 150Hz for the setup in Figure 3a and $f = f_{clock}$ + 150Hz for the setup in figure S4.

## Authors contributions

D.P. developed the theoretical analysis of signal-to-noise detection performances into a heterodyne detection system with commercial mid-infrared detectors. D.P., L.M., P.U., M.P., F.C., P.D.N. and S.B. conceived the experiments and contributed with technical insights for data acquisition. L.M. and D.P. carried out the measurements, with the support of C.V., P.U., M.P., C.L., F.C. and S.B. D.P. wrote the initial version of the manuscript, with the support of L.M. All authors contributed to revise the manuscript.

## Acknowledgments

This project has received funding from the European Defence Fund (EDF) under grant agreement 101103417 EDF-2021-DIS-RDIS-ADEQUADE. D.P., L.M., C.V. and L.D. acknowledge funding from the European Union's Horizon Europe research and innovation program under the project "INPHOMIR" (grant agreement No. 101135749 HORIZON-CL4-2023-DIGITAL-EMERGING-01). F.C., P.D.N. and S.B. acknowledge funding from the European Union's NextGenerationEU Programme with the I-PHOQS Infrastructure [IR0000016, ID D2B8D520] "Integrated infrastructure initiative in Photonic and Quantum Sciences".

Funded by the European Union. Views and opinions expressed are however those of the author(s) only and do not necessarily reflect those of the European Union or the European Commission or the European Health and Digital Executive Agency (HADEA). Neither the European Union nor the granting authority can be held responsible for them.

## Competing interests

The authors declare no competing interests

# Supplementary Materials

1. **Analysis of current noise terms in MIR photo-detectors employed in a heterodyne scheme**

As described in the main text, the current noise in photodetectors employed in a heterodyne scheme can be divided in two groups $S_{det}$ (noise terms that are intrinsic to the detectors) and $S_{LO}$ (noise terms that are induced by the LO). The detector noise can be expressed as:

(S1.1)  $S_{det} = S_{d+bg} + S_{th} + S_{1/f} + S_{TIA}$

Where the first term is the dark current noise $i_d$ and the background noise $i_{bg}$ (photocurrent generated by the room-temperature background):

(S1.2)  $S_{d+bg} = 2eg(i_d + i_{bg})$;

The second term is the Johnson-Nyquist noise, i.e. the noise generated by the thermal fluctuations and is proportional to the Boltzman constant $k_b$, the detector temperature T and its differential resistance $r$:

(S1.3)  $S_{th} = 4k_b T/r$

The last two terms of (S1.1) are pure electronic terms, the TIA thermal noise $S_{TIA} = 4k_b T/r_{TIA}$ and the 1/f noise. **Figures S1** shows the intrinsic detector current noise for the QCD and the MCT photodetectors employed in this work, obtained by combing the parameters of **Table 1** and a simple model based on equation 2: we can notice that detector noise is flat between 100 kHz and hundreds of MHz for the MCT and up to tens of GHz for the QWIP and the QCD devices.

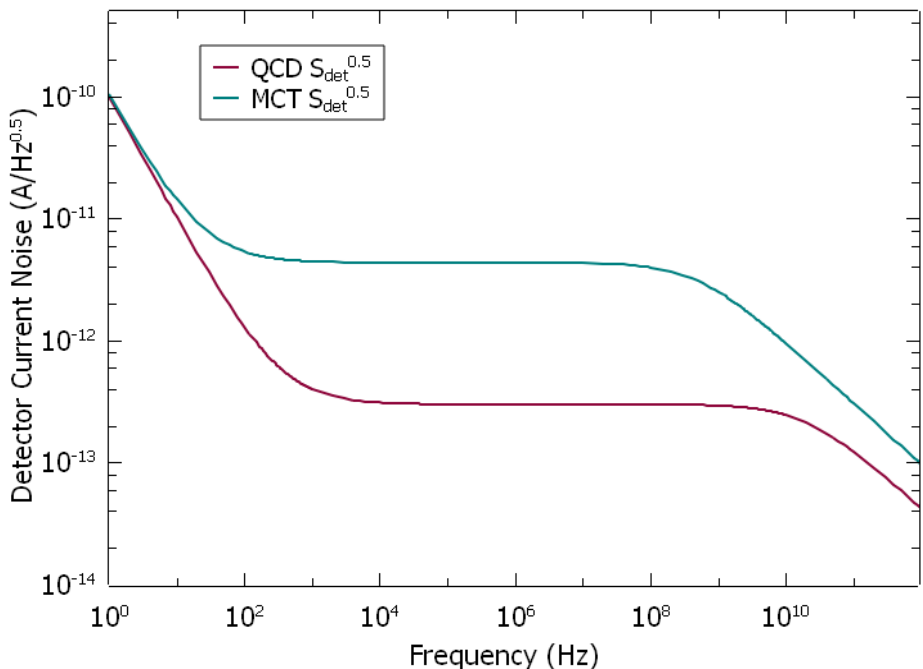

**Figure S1** Intrinsic spectral current noise for QCD and MCT photodetectors

As in this work we are interested in the strength of the signal rather than the large detector bandwidth, we have performed heterodyne detection measurements in the accessible range 70-150 MHz: at these frequencies we are far from the 1/f noise and, regarding the relative intensity noise (RIN), typical RIN values of QCLs at 4-5µm measured at f ≥ 50MHz for ~10mW output power are in the range [-150dB/Hz : -180dB/Hz][1] corresponding to $(S_{RIN})^{1/2}$ <1.0x10$^{-15}$A/Hz$^{1/2}$ (which is negligible as compared to $S_{det}$ in **Fig. S1).**

The laser-induced noise term can be expressed as:

(S1.4)  $S_{LO} = S_{shot} + S_{LFN} + S_{RIN}$

Where the first term is the poissonian shot noise:

(S1.5)  $S_{shot} = 2egI_{LO} = 2egRP_{LO}$

And the last two terms of (S1.4) are the laser-frequency-noise $S_{LFN}$ [2] which is proportional to the current-driver noise $S_{driver}(f)$ and the relative-intensity-noise $S_{RIN}$ that is inversely proportional to the LO power. The table below compares the quantities of all these current noise terms.

| Employed Photodetector | MCT | QCD |
|---|---|---|
| Heterodyne parameters | $P_{LO}$ = 1 mW, f = 100MHz | $P_{LO}$ = 30 mW, f = 100MHz |
| Noise current term [A/Hz$^{0.5}$] | | |
| $\sqrt{S_{d+bg}}$ | 8.5x10$^{-13}$ | 1.5x10$^{-14}$ |
| $\sqrt{S_{th}}$ | 3.5x10$^{-12}$ | 4.2x10$^{-13}$ |
| $\sqrt{S_{1/f}}$ | ~1.0x10$^{-18}$ | ~1.0x10$^{-19}$ |
| $\sqrt{S_{TIA}}$ | 1.1x10$^{-12}$ | - |
| $\sqrt{S_{shot}}$ | 2.1x10$^{-11}$ | 4.4x10$^{-13}$ |
| $\sqrt{S_{LFN}}$ | ~1.0x10$^{-13}$ | ~1.0x10$^{-15}$ |
| $\sqrt{S_{RIN}}$ | ~1.0x10$^{-14}$ | ~1.0x10$^{-15}$ |

**Figure S2** illustrates the current noise of the detectors employed in a heterodyne setup, as function of the Local Oscillator Power.

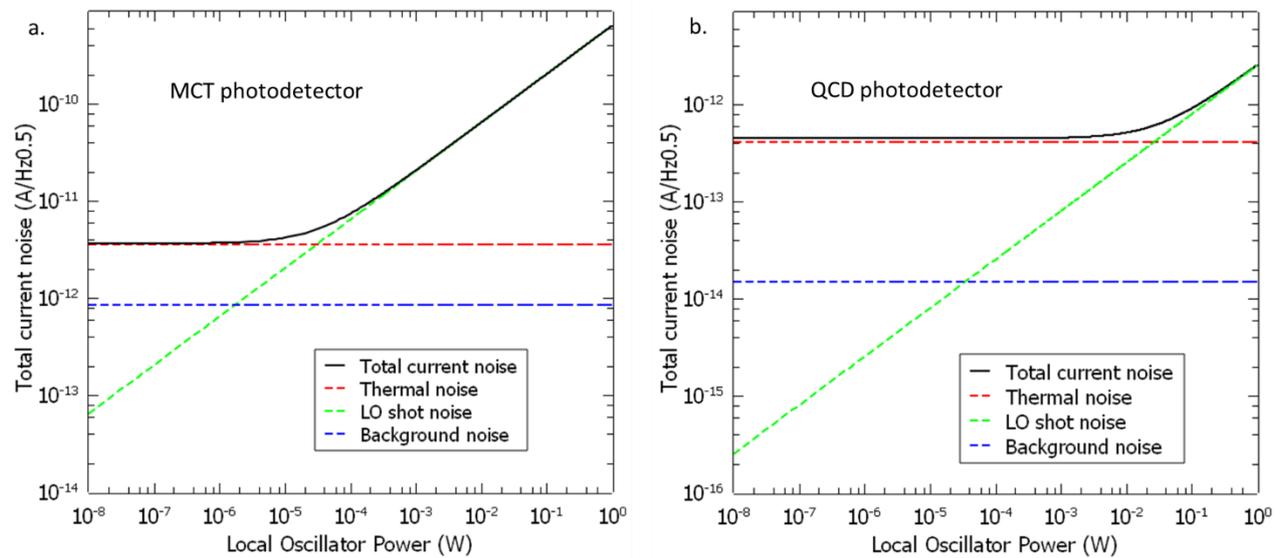

**Figure S2** Total current noise as function of the Local Oscillator Power for MCT (a) and QCD (b) photodetectors.

## 2. Free-running SHD and BHD setups with MCT and QCD detectors

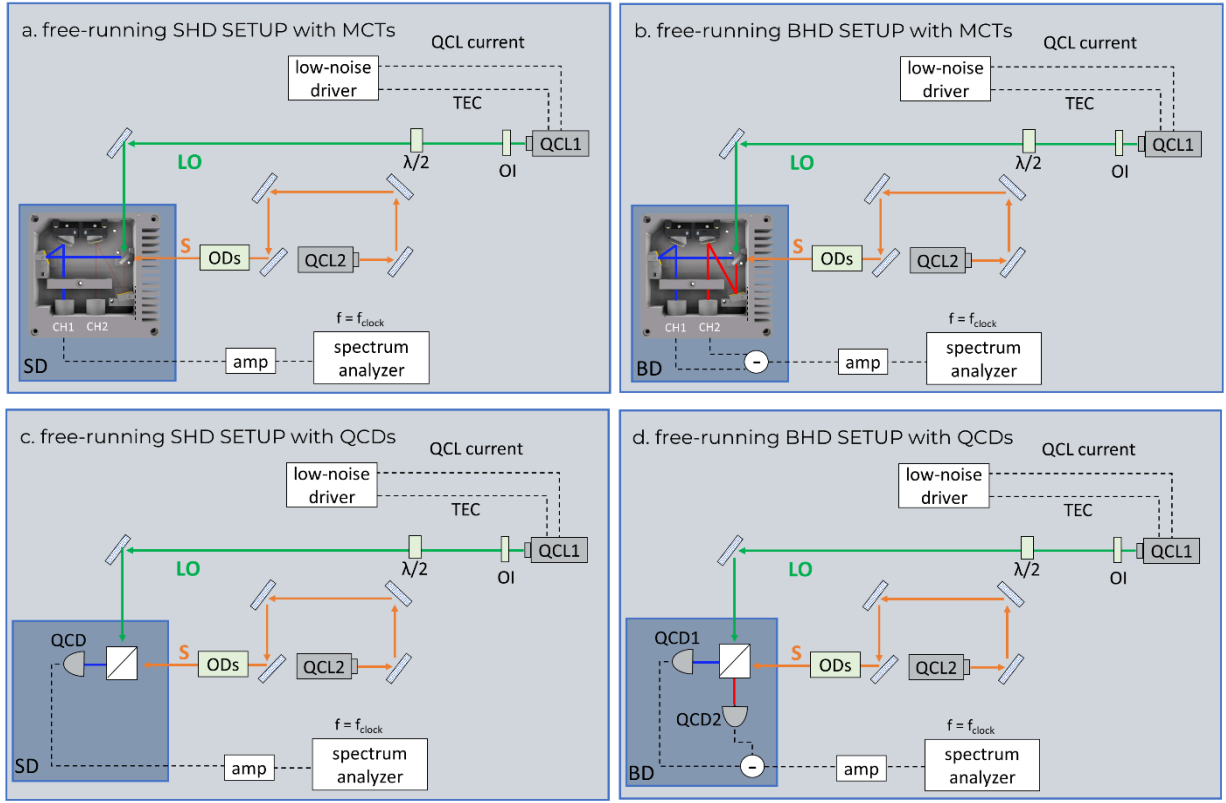

**Figure S3** Setups employed for free-running SHD and free-running BHD acquisitions.

## 3. BHD+PLL setup with a pair of QCDs

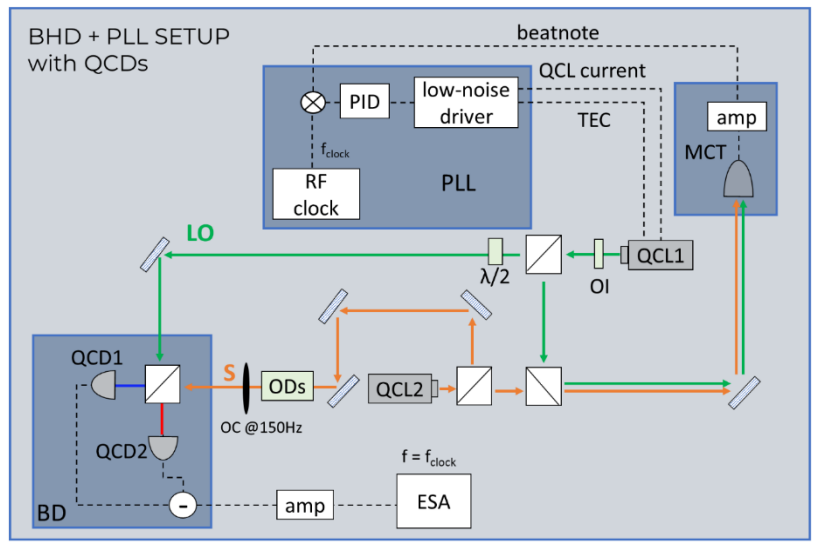

**Figure S4** BHD+PLL setup with QCDs and a beam-splitter in calcium fluoride.

## 4. Heterodyne spectra

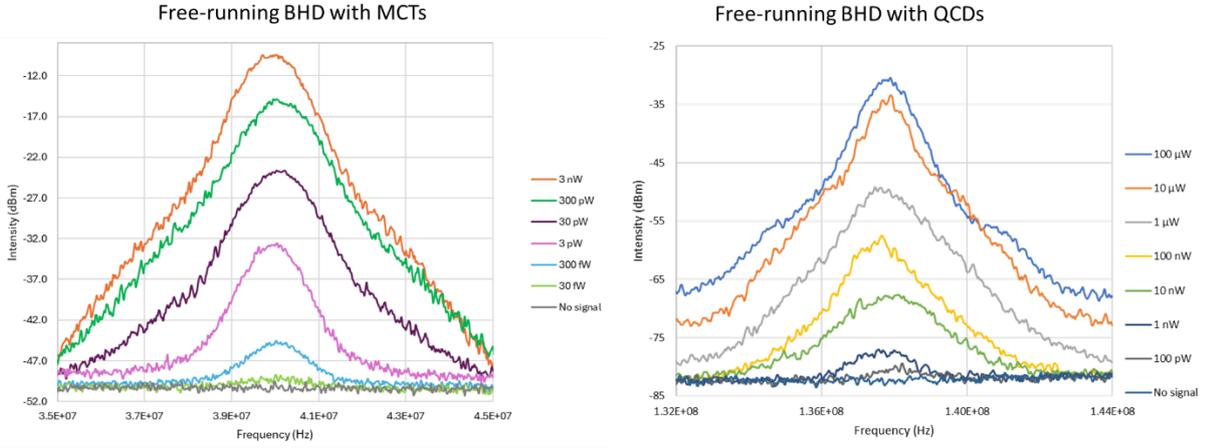

**Figure S5** Heterodyne spectra obtained with free-running BHD setups employing MCT (left) and QCD (right) detectors.

## 5. Heterodyne interferometry

In a classic Mach-Zender Interferometer (MZI), the combination of the two identical fields $E_s$ shifted by a phase $\Delta\Phi$ generates an output at the photodiode which is expressed as:

(S4.1) $\quad \left|E_s e^{i\omega_s t} + E_s e^{i\omega_s t + i\Delta\Phi}\right|^2 = \left|E_s e^{i\omega_s t}(1 + e^{i\Delta\Phi})\right|^2 = 2|E_s|^2 (1 + \cos(\Delta\Phi))$

If we combine a MZI signal with a BHD setup, we produce an interferometric signal at the beatnote of $E_s$ and $E_{LO}$:

(S4.2) $\quad \left|E_{LO} e^{i\omega_{LO}} + E_s e^{i\omega_s t}(1 + e^{i\Delta\Phi})\right|^2 = 2|E_s E_{LO}| \cdot \cos[(\omega_S - \omega_{LO})t] \cdot (1 + \cos(\Delta\Phi))$

The Heterodyne interferometry photocurrent can therefore be expressed as:

(S4.3) $\quad i_{het} = 2R\sqrt{P_S P_{LO}} \cdot \cos[(\omega_S - \omega_{LO})t] \cdot (1 + \cos(\Delta\Phi))$

Finally, by computing the heterodyne intensity signal we obtain the phase dependency $(1 + \cos(\Delta\Phi))^2$, as it is observed in the experimental data in Figure 4c.

(S4.3) $\quad I_{het} = 4R^2 Z_{50\Omega} P_S P_{LO} \cdot (1 + \cos(\Delta\Phi))^2 \cdot \cos[(\omega_S - \omega_{LO})t]$